\documentclass[12pt,titlepage]{article}
\usepackage[letterpaper,top=1in,bottom=1in,left=1.25in,right=1.25in]{geometry}
\usepackage{graphicx,cite, url}

\usepackage{mathrsfs,cleveref,amsmath}

\newcommand{\ft}{\widetilde}
\newcommand{\mr}{\mathrm}
\newcommand{\Deps}{\Delta\epsilon}
\newcommand{\IFT}{\mathscr{F}^{-1}}
\newcommand{\tf}{\mathcal{H}}

\begin{document}

\title{High-throughput intensity diffraction tomography with a computational microscope}

\author{Ruilong Ling$^{1,\dagger}$, Waleed Tahir$^{1,\dagger}$, Hsing-Ying Lin$^{2,3}$, Hakho Lee$^{2,3}$, Lei Tian$^{1,*}$
\\
\multicolumn{1}{p{\textwidth}}{\centering\emph{\normalsize 1. Department of Electrical and Computer Engineering, Boston University, Boston, MA 02215, USA\\
		2. Center for Systems Biology, Massachusetts General Hospital, Boston, MA 02114, USA\\
		3. Department of Radiology, Massachusetts General Hospital, Harvard Medical School, Boston, MA 02114, USA\\
		$^{\dagger}$ These authors contributed equally to this work\\
		$^{*}$ lei\_tian@alum.mit.edu
}}}

\maketitle

\begin{abstract}
	We demonstrate a motion-free intensity diffraction tomography technique that enables direct inversion of 3D phase and absorption from intensity-only measurements for weakly scattering samples.   We derive a novel linear forward model, featuring slice-wise phase and absorption transfer functions using angled illumination.  This new framework facilitates flexible and efficient data acquisition, enabling arbitrary sampling of the illumination angles.   The reconstruction algorithm performs 3D synthetic aperture using a robust, computation and memory efficient  slice-wise deconvolution to achieve resolution up to the incoherent limit.    We demonstrate our technique with thick biological samples having both sparse 3D structures and dense cell clusters.  We further investigate the limitation of our technique when imaging strongly scattering samples.  Imaging performance and the influence of multiple scattering is evaluated using a 3D sample consisting of stacked phase and absorption resolution targets.   This computational microscopy system is directly built on a standard commercial microscope with a simple LED array source add-on, and promises broad applications by leveraging the ubiquitous microscopy platforms with minimal hardware modifications.
\end{abstract}


	
	\section{Introduction}
	
	Quantitative characterization of thick biological samples is a challenging task. Unstained biological cells appear  transparent when imaged under a standard brightfield microscope, since only the sample's absorption information is directly visible.  Though techniques based on exogenous labels (e.g. dyes and fluorophores) have been developed~\cite{Stephens2003}, they suffer from the need for staining or labeling using external contrast agents which may alter cellular behavior~\cite{Waeldchen2015, Hoebe2007}.  Here, we develop a new {\it label-free} phase tomography technique, which provides 3D cellular information with intrinsic structural sensitivity.  Our technique is fast, motion-free, and easy to implement with a computational microscope platform, in which a standard commercial microscope is modified with an LED array source~\cite{Zheng2011, Zheng2013, Tian2014, Tian2014a, Tian.Waller2015, horstmeyer2016diffraction, Tian2015a, Tian2015b}. 
	
	3D phase microscopy techniques can be largely categorized into two classes, interferometry based and intensity-only methods.  The most widely used {\it interferometry-based} technique is optical diffraction tomography (ODT).   In ODT,  images are first taken {\it interferometrically} to directly record both phase and amplitude information of the scattered field.  Next, a tomographic reconstruction algorithm is devised to recover the sample's 3D refractive index distribution.  The interferograms are taken by using either a separate reference path~\cite{Choi.etal2007, Sung.etal2009, cotte2013marker, kamilov2015learning, simon2017tomographic, Charriere:06, Zhang:2016aa} or a common-path interferometer attached to an existing tomography setup~\cite{Bon2014a, Kim.etal2014, nguyen2017gradient, kim2014common}.  Various tomographic measurement schemes have been developed, including projection measurement by rotating the sample mechanically~\cite{Charriere:06, simon2017tomographic} or with an optical tweezer~\cite{kus2014tomographic, habaza2015tomographic, kim2015simultaneous, kim2017tomographic}, varying the illumination angles with a tilting mirror~\cite{Choi.etal2007, Sung.etal2009, cotte2013marker, kim2014common, kamilov2015learning, simon2017tomographic} or a spatial light modulator~\cite{shin2015active}, and through-focus measurement with a mechanical stage~\cite{Bon2014a, Kim.etal2014, nguyen2017gradient}.  Due to the need for interferometry, ODT typically requires additional specialized and expensive hardware, which is not always compatible to standard microscopes.  Naturally, one would prefer a technique capable of leveraging the ubiquitous microscopy platforms with minimal hardware modifications. To this end, there has been a continued interest in {\it intensity-only} phase tomography techniques, which perform 3D phase imaging without interferometry\cite{Maleki:93, Gbur:02ol, Gbur:02, Gbur:05, Anastasio:05, Jenkins2015, Soto2017, Rodrigo2017, soto2018optical, Chen2016, Tian2014a, Tian.Waller2015, horstmeyer2016diffraction}. 
	
	{\it Intensity} diffraction tomography (IDT) refers to a class of 3D imaging techniques that employ tomographic phase reconstruction from {\it intensity-only} measurements.    One IDT approach~\cite{Anastasio:05, Gbur:02ol, Gbur:02, Gbur:05}  combines a defocus-based phase contrast technique~\cite{N19846} and diffraction tomography model~\cite{Wolf1969} to recover 3D phase.  The measurement involves taking multiple defocused images while rotating the sample or the illumination.  Recent works further incorporate partially coherent illumination using symmetric~\cite{Jenkins2015, Soto2017, Rodrigo2017, soto2018optical} or asymmetric~\cite{Chen2016} light source to achieve up to 2$\times$ resolution improvement in the recovered phase.   However, changing focus not only requires mechanical scanning, but also increases the acquisition time and data size, both of which are undesirable for high-throughput applications.   
	
	An alternative IDT approach extracts 3D phase information using angled illumination without mechanical scanning~\cite{Tian2014a, Tian.Waller2015, horstmeyer2016diffraction}.  In \cite{Tian2014a}, 3D phase contrast is computed with an algorithm  combining lightfield refocusing~\cite{Levoy2006} and differential phase contrast~\cite{Mehta2009}; however, the results suffer from low-resolution as diffraction effects are neglected.  In~\cite{Tian.Waller2015}, a multislice model is proposed to incorporate diffraction and multiple forward scattering, and achieves high-resolution  3D recovery.  However, since the multislice model is {\it nonlinear}, it necessitates an {\it iterative} reconstruction algorithm, which is non-ideal for time-constrained applications.   In \cite{horstmeyer2016diffraction}, an {\it iterative} algorithm combining  Fourier ptychography~\cite{Zheng2013} and the first Born approximation~\cite{Wolf1969} was proposed.  However, their model ignored the interference term between the scattered and unscattered fields. Here, we show that this term is the {\it primary} source of  3D phase contrast in IDT measurements.
	
\begin{figure}[t]
		\centering
		\includegraphics[width=\linewidth]{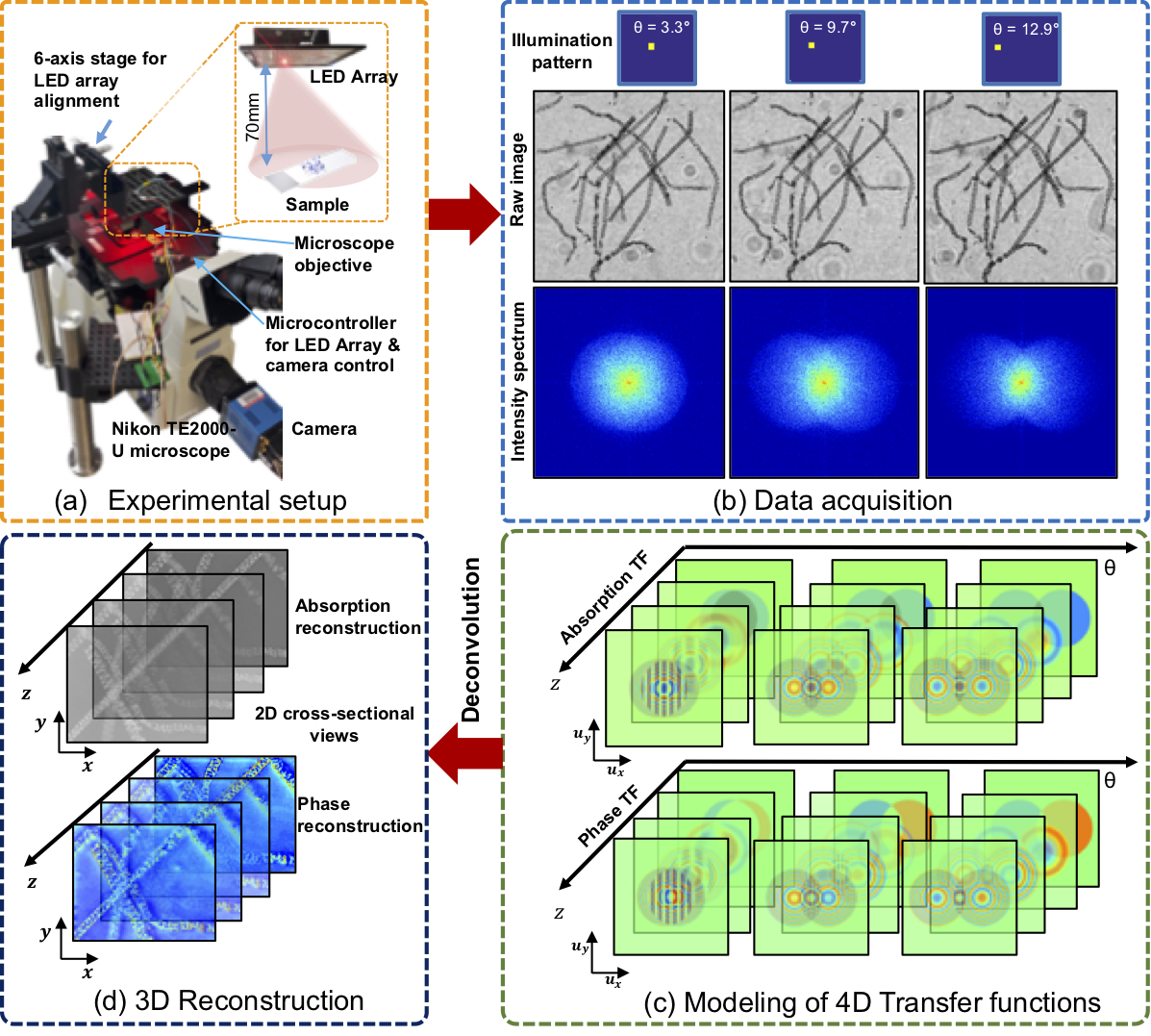}
		\caption{Intensity diffraction tomography from angled illumination. (a) The setup consists of a standard microscope with an LED array that allows flexible patterning of illumination angles. (b) Images are taken by varying the illumination angle. Each intensity spectrum of the raw data exhibits two shifted circles, whose shift is set by the illumination angle. (c) Corresponding phase (imaginary part) and amplitude (real part) transfer functions (TF) for the same set of illumination angles are visualized at various sample depths. (d) The slice-wise deconvolution algorithm outputs two 3D stacks, corresponding to the phase and absorption reconstruction. }
		\label{fig:setup}
	\end{figure}

	In this work, we develop a novel {\it linear} IDT model that relates the sample's 3D permittivity contrast to  intensity measurements using angled illumination (Fig.~\ref{fig:setup}).   Previous efforts~\cite{Jenkins2015, Soto2017, Rodrigo2017, Chen2016, soto2018optical} formulate the phase-intensity mapping in the {\it 3D} Fourier space;  this approach unfortunately suffers from stringent sampling requirement for the measurement, resulting in hundreds of defocused images needed in practice.   In addition, the reconstruction requires computation and memory intensive 3D deconvolution.   Our approach overcomes {\it all} these limitations by employing a slice-based framework.  The 3D sample is first modeled as a series of 2D slices along the axial direction.    We then derive the slice-wise (2D) phase and absorption transfer functions (TF) at different depths for each illumination angle.    We show that this framework enables flexible and efficient data acquisition, allowing using arbitrary patterning of the illumination angles and much fewer images required compared to other techniques~\cite{Jenkins2015, Soto2017, Rodrigo2017, Chen2016, soto2018optical, horstmeyer2016diffraction}.   Our model fully accounts for the interference between the scattered and unscattered fields.    The linearization is achieved via the first Born approximation that considers single scattering and neglects higher order nonlinear effects.    Our linear model enables {\it non-iterative} 3D reconstruction directly from intensity-only measurements.  We demonstrate volumetric reconstructions with closed-form Tikhonov-regularized solutions, implemented using a computation and memory efficient 2D FFT-based algorithm.  We demonstrate our technique on both stained and unstained thick biological samples.  The imaging performance and limitation is further investigated in the presence of strong multiple scattering.  Experiments demonstrate that our technique is robust even for samples with large permittivity contrast.

	\section{Theory and method}
	\label{sec:method}
	
	\subsection{Forward model}
	
	The scattering of a sample can be characterized by its scattering potential $V(\vec{r}) = \frac{1}{4\pi}k_0^2\Deps(\vec{r})$~\cite{Born1999},  where $\Deps=\Deps_{\mr{Re}} + \mr{i} \Deps_{\mr{Im}} = \epsilon - \epsilon_0$ is the permittivity contrast between the sample $\epsilon$ and the surrounding medium $\epsilon_0$. The real part $\Deps_{\mr{Re}}$ characterizes the phase effect, and the imaginary part $\Deps_{\mr{Im}}$ describes absorption. $k_0 = 2\pi/\lambda$ is the wave number in free space, where $\lambda$ the wavelength of the illumination. $\vec{r} = (\vec{x},z)$ denotes the 3D spatial coordinates, with transverse coordinates $\vec{x}$ and axial position $z$. 
	
	We employ the first Born approximation to model the light-sample interaction. Given the incident field $f_i$, the total field $f$ after propagating through the sample is given by
	\begin{equation}
	f(\vec{r}) = f_{i}(\vec{r}) + \int f_{i}(\vec{r'}) V(\vec{r'}) G(\vec{r} - \vec{r'})\mr{d}^3\vec{r'},
	\label{e1}
	\end{equation}
	where $G(\vec{r}) = \exp(\mr{i}k|\vec{r}|)/|\vec{r}|$ is the outgoing Green's function and $k=\sqrt{\epsilon_0}k_0$.  The integral term represents the field $f_s$ scattered off the sample.    The plane-wave illumination is modeled as $f_{i}(\vec{r}| \vec{u}_i) = \sqrt{S(\vec{u}_i)}e^{-\mr{i}(\vec{u}_i\cdot\vec{x}+\eta_i z)}$ with transverse frequency $\vec{u}_i$ and axial frequency $\eta_i = \sqrt{k^2-|\vec{u}_i|^2}$. $S(\vec{u}_i)$ represents the intensity of the $i^{th}$ LED, which we model as a point source~\cite{Zheng2011, Zheng2013, Tian2014, Tian2014a, Tian.Waller2015, horstmeyer2016diffraction, Tian2015a, Tian2015b}.  The notation $a (\cdot | \vec{u_i}) $ indicates the illumination direction for the field $a$.  As the field propagates through the microscope, it is filtered by the pupil $P(\vec{u})$, corresponding to a convolution with the coherent point spread function $h(\vec{x})$.   The resulting intensity at the back focal plane of the microscope $I(\vec{x},0| \vec{u}_i)$ is
	\begin{equation}
	I(\vec{x},0| \vec{u}_i) =  \left|f(\vec{x},0| \vec{u}_i)\ast h(\vec{x})\right|^2, 
	\label{e3}
	\end{equation}
	where $\ast$ denotes 2D convolution.  For notational simplicity, we have neglected the microscope's magnification.  The total field at the front focal plane ($z = 0$) is $f(\vec{x},0| \vec{u}_i) = f_{i}(\vec{x},0| \vec{u}_i) + f_s(\vec{x},0| \vec{u}_i)$, in which the scattered field is simplified using the Weyl expansion of the Green's function~\cite{Born1999} as
	\begin{equation}
	f_s(\vec{x},0| \vec{u}_i) = 
	{\frac{\mr{i}k_0^2}{2}\sqrt{S(\vec{u}_i)}}  
	\int
	\IFT \left\{ \ft{\Deps}(\vec{u}+\vec{u}_i,z')\frac{e^{-\mr{i}(\eta(\vec{u})+\eta_i) z'}}{\eta(\vec{u})} \right\}\mr{d}z',
	\label{e2}
	\end{equation}
	where $\ft{\cdot}$ represents the 2D Fourier transform (FT);  $\ft{\Deps}(\vec{u},z)$ is the 2D FT of the slice $\Deps(\vec{x},z)$ at depth $z$; $\IFT\{\cdot\}$ denotes the 2D inverse FT (IFT); $\vec{u}$ is the transverse frequency variable, and $\eta(\vec{u})=\sqrt{k^2-|\vec{u}|^2}$ the axial frequency.  \eqref{e2} calculates the total scattered field as the coherent superposition of the scattered field from each slice; inter-slice scattering is ignored as a consequence of the first Born approximation. 
	
	Importantly, \eqref{e3} contains the interference information between the unscattered and scattered field, and can be expanded into four terms: $I = I_i + I_{ss} + I_{is} + I_{si}$. $I_i=  S(\vec{u}_i)|P(-\vec{u}_i)|^2$ is the constant background intensity. $I_{ss} =  |f_s(\vec{x},0| \vec{u}_i)\ast h(\vec{x})|^2$ represents squared modulus of the scattered field, which is negligible if the permittivity contrast is small, and can thus be dropped~\cite{Streibl:85}.  Hence, the phase and absorption information is mainly contained only in the two cross-terms, $I_{is} =  [f_{i}(\vec{x},0| \vec{u}_i)\ast h(\vec{x})]^{*}[f_s(\vec{x},0| \vec{u}_i)\ast h(\vec{x})]$ and $I_{si} = I^*_{is}$.  This is also highlighted by the Fourier spectrum of the intensity measurement in Fig. \ref{fig:setup}.  The relation between the sample's permittivity contrast and the intensity spectrum is thus {\it linear}; and the phase and absorption terms are {\it decoupled}:
	\begin{multline}
	\ft{I} (\vec{u},0| \vec{u}_i)
	\approx 
	S(\vec{u}_i)|P(-\vec{u}_i)|^2\delta(\vec{u}) +\\
	\int \left[H_{\mr{Re}}(\vec{u}, z |\vec{u}_i)\ft{\Deps}_{\mr{Re}}(\vec{u},z)
	+H_{\mr{Im}}(\vec{u}, z |\vec{u}_i)\ft{\Deps}_{\mr{Im}}(\vec{u},z)\right]
	\mr{d}z  ,
	\label{e6}
	\end{multline}
	where $H_{\mr{Re}}$ and $H_{\mr{Im}}$ are the {\it angle-dependent} phase and absorption TFs for {\it each sample slice} at depth $z$, respectively, and are
	\begin{align}
	H_{\mr{Re}}(\vec{u}, z |\vec{u}_i) &= 
	\frac{\mr{i}k_0^2}{2}S(\vec{u}_i) 
	\Bigg \{P^{*}(-\vec{u}_i)
	\frac{e^{-\mr{i}[\eta_i+\eta(\vec{u}-\vec{u}_i)] z}}{\eta(\vec{u}-\vec{u}_i)}
	P(\vec{u}-\vec{u}_i)   \nonumber\\
	& - P(-\vec{u}_i)
	\frac{e^{\mr{i}[\eta_i+\eta(\vec{u}+\vec{u}_i)] z}}{\eta(\vec{u}+\vec{u}_i)}
	P^{*}(-\vec{u}-\vec{u}_i)\Bigg \},\\
	H_{\mr{Im}}(\vec{u},z |\vec{u}_i) &=
	-\frac{k_0^2}{2}S(\vec{u}_i)
	\bigg\{P^*(-\vec{u}_i)
	\frac{e^{-\mr{i}[\eta_i+\eta(\vec{u}-\vec{u}_i)] z}}{\eta(\vec{u}-\vec{u}_i)}
	P(\vec{u}-\vec{u}_i)     \nonumber\\
	& +P(-\vec{u}_i)
	\frac{e^{\mr{i}[\eta_i+\eta(\vec{u}+\vec{u}_i)] z}}{\eta(\vec{u}+\vec{u}_i)}
	P^{*}(-\vec{u}-\vec{u}_i)\bigg \}.
	\label{e7}
	\end{align}
	
	Equations~(\ref{e6}-\ref{e7}) define the forward model of our technique.   From these equations, the computational complexity of our model is set by the $N_z$ evaluations of 2D FFT or $O(N_xN_yN_zlog(N_xN_y))$, where $N_x,N_y,N_z$ correspond to the dimension of the data in 3D. The TFs have the following essential properties for scan-free 3D reconstruction.  {\bf (a)} The transverse frequency of the incident field $\vec{u_i}$ determines the off-axis shift of the pupil function.   Due to {\it intensity-only} measurement, a pair of shifted pupils, shifting to opposite directions,  are super-imposed in the TFs (akin to the ``twin-image'' in holography).  This is illustrated in the computed phase and absorption TFs in Fig.~\ref{fig:setup}(c), and validated {\it experimentally} by visualizing the intensity spectrum of the raw images in Fig.~\ref{fig:setup}(b).  {\bf (b)} The depth information is encoded in the phase term, which enables {\it scan-free} 3D reconstruction.  Specifically, the linear phase term $e^{\mr{i}\eta_i z}$ corresponds to a geometrical shift in the real space, which was accounted for previously using the lightfield model~\cite{Tian2014a}; the propagation term $e^{\mr{i}\eta(\vec{u}+\vec{u}_i)z}$ models the diffraction effects.  {\bf (c)} At the focal plane, assuming a real and symmetric pupil function (i.e. an ideal unaberrated microscope), the {\it phase} TF is imaginary and anti-symmetric, whereas the {\it absorption} TF is real and symmetric.  In general, because the phase and absorption information are encoded {\it asymmetrically} in the Fourier space [Fig.~\ref{fig:setup}(c)], it leads to measurable intensity contrast difference for phase and absorption features, and allows inversion of both quantities simultaneously.  {\bf (d)} The support of the combined TFs sets the resolution limit of our technique, which can be analyzed following the same framework as~\cite{Jenkins2015, Soto2017, Rodrigo2017, Chen2016, soto2018optical}.   Since the forward model relies on the existence of a strongly unscattered field, it is only valid for brightfield measurements.  As a result, we use illumination angles up to the objective NA.   This allows us to achieve resolution equivalent to the {\it incoherent} limit.  In the transverse direction, the support is uniformly ${4\mr{NA}}/{\lambda}$.  In the axial direction, the system suffers from the missing cone problem~\cite{Streibl:85, Sheppard1994}, resulting in the axial elongation present in the reconstruction.  The axial Fourier coverage varies with sample's feature size, and is up to $\left({2-2\sqrt{1-\mr{NA}^2}}\right)/{\lambda}$.

	\begin{figure*}[!h]
		\centering
		\includegraphics[width=\linewidth]{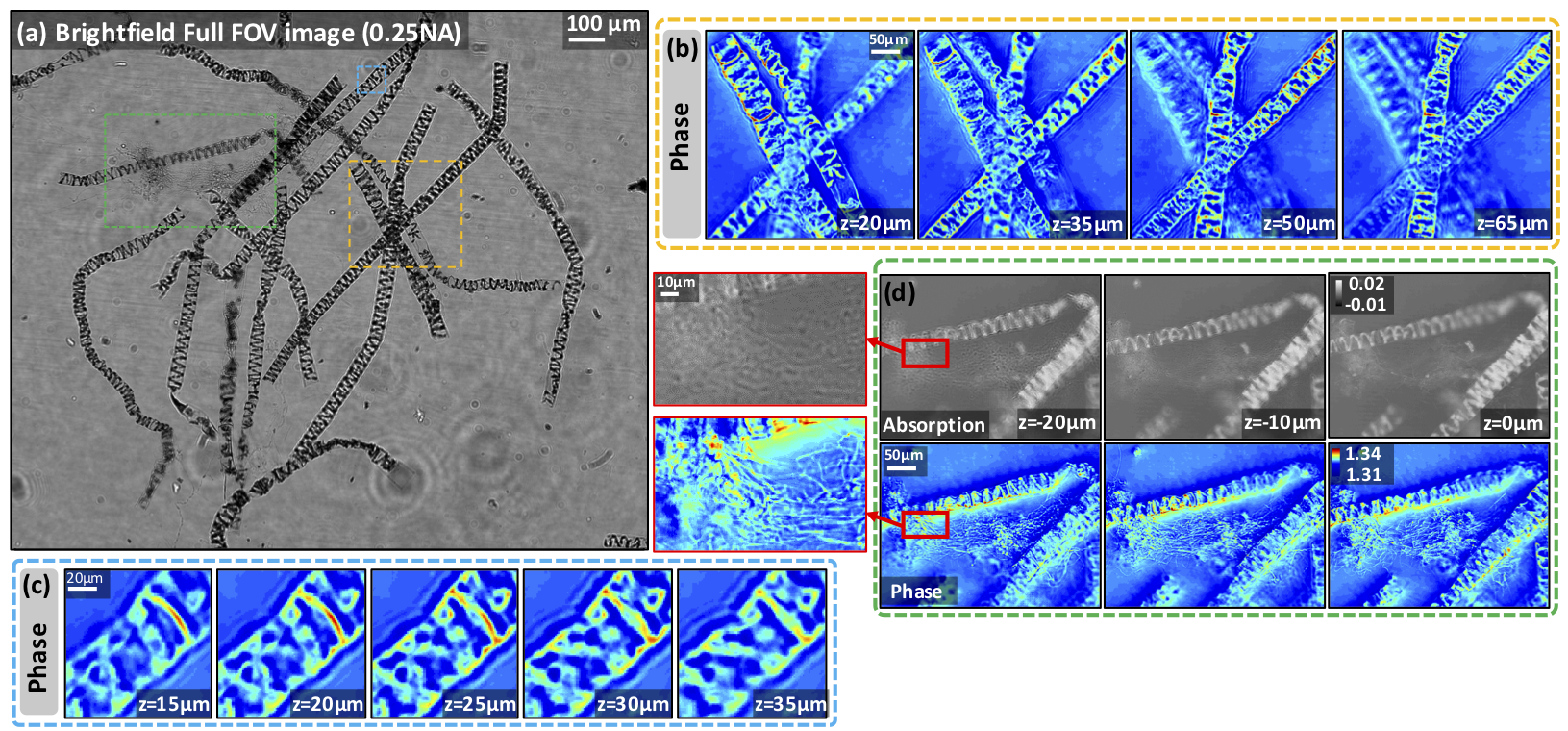}
		\caption{Phase and absorption reconstruction of a stained spirogyra sample. (a) The full field of view (FOV) brightfield image with the on-axis LED illumination (10$\times$, 0.25NA).  The sample contains both highly absorbing features (e.g. chloroplasts)  and ``phase'' features (e.g. filaments).  (b) A dense algae cluster is successfully resolved in the phase reconstruction.  (c) Phase reconstruction of spiral structures on a single spirogyra, further demonstrating the axial sectioning capability of our technique.  (d) Unstained filaments are resolved with high contrast in the phase reconstruction whereas the absorption does not provide much contrast. 	}
		\label{spiral}
	\end{figure*}
	
	\subsection{Inverse problem}
	
	The reconstruction algorithm combines all the intensity images to estimate the sample's complex permittivity contrast (i.e. phase and absorption) in 3D.   Since our forward model enables inferring the full field information, the reconstruction essentially stitches all the Fourier components {\it coherently}, akin to the 3D synthetic aperture~\cite{lue2008synthetic, lee2013synthetic}.   To facilitate efficient reconstruction, we propose a {\it slice-wise} deconvolution algorithm, in which the phase and absorption are reconstructed slice-by-slice.      
	
	To implement our algorithm, we first discretize the 3D sample into a stack of 2D slices, equivalently replacing the integral in \eqref{e6} by a discrete sum over the slice index.  Next, each intensity image is pre-processed to perform background normalization and removal: $g = (I-I_i)/I_i$.  The corresponding discretized and normalized TFs  are: $\tf_{\mr{Re}}=H_{\mr{Re}}/I_i$ and $\tf_{\mr{Im}}=H_{\mr{Im}}/I_i$.  This normalization also removes the unknown scaling in the source intensity.    Direct deconvolution is known to suffer from noise amplifications~\cite{bertero1998introduction}; therefore we employ Tikhonov regularization that imposes an energy constraint to suppress these artifacts.  The {\it closed-form} solutions for phase and absorption are 
	\begin{align}
	& \Deps_{\mr{Re}}[m]  = 
	\IFT
	\Bigg \{ \frac{1}{A}
	\bigg \{\big(\sum_l \big|\tf_{\mr{Im}}[l,m]\big|^2+\beta\big)
	\odot \big(\sum_l\tf^\ast_{\mr{Re}}[l,m]\odot \ft{g}[l] \big)  \nonumber\\
	& - \big(\sum_l\tf^\ast_{\mr{Re}}[l,m] \odot \tf_{\mr{Im}}[l,m] \big)
	\odot \big( \sum_l\tf^\ast_{\mr{Im}}[l,m]\odot \ft{g}[l]  \big)
	\bigg \}
	\Bigg \},   \label{e9}\\ 
	& \Deps_{\mr{Im}}[m]  = 
	\IFT
	\Bigg \{ \frac{1}{A}
	\bigg \{ \big(\sum_l \big|\tf_{\mr{Re}}[l,m] \big|^2+\alpha\big)
	\odot \big(\sum_l\tf^\ast_{\mr{Im}}[l,m]\odot \ft{g}[l] \big)   \nonumber\\
	& - \big(\sum_l\tf_{\mr{Re}}[l,m]\odot \tf^\ast_{\mr{Im}}[l,m]\big)
	\odot \big(\sum_l\tf^\ast_{\mr{Re}}[l,m]\odot \ft{g}[l]\big)
	\bigg \} 
	\Bigg \}, \label{e10}
	\end{align}
	where $A  = \big(\sum_l|\tf_{\mr{Re},l}|^2 + \alpha\big)\odot
	\big(\sum_l|\tf_{\mr{Im},l}|^2 + \beta\big)   
	- \big(\sum_l\tf_{\mr{Re},l}\odot \tf^\ast_{\mr{Im},l}\big)\odot 
	\big(\sum_l\tf^\ast_{\mr{Re},l}\odot \tf_{\mr{Im},l}\big)$; $[m]$ indexes the $m$th  sample slice, $[l]$ the $l$th intensity image, and $[l,m]$ the TFs for the $m$th slice from the $l$th illumination; $\odot$ denotes element-wise multiplication between two matrices, $\alpha$ and $\beta$ are the regularization parameters for phase and absorption, respectively.

	\section{Results}
	\label{sec:result}

	Our system consists of a Nikon TE 2000-U microscope with a programmable red (central wavelength 630nm) LED array source (specifications same as ~\cite{Tian2015b}) placed 79mm above the sample [Fig.~\ref{fig:setup}(a)].  An sCMOS camera (Pco.Edge 5.5) is used for image acquisition, which is synchronized in real-time with the LED source via a microcontroller.  Since our technique completely removes any mechanical scanning, the acquisition speed is only limited by the camera's frame rate (up to 50Hz).  To achieve the incoherent resolution limit, we acquire data using brightfield LEDs fully covering the NA of the objective used in each experiment.  Angle-varying intensity images are captured by sequentially turning on one LED at a time.   We conduct experiments using 10$\times$ (0.25 NA, CFI Plan Achro) and 40$\times$ (0.65 NA, CFI Plan Achro) microscope objectives (MO).   For the 10$\times$ MO, 89 LEDs are within the brightfield region; whereas 697 LEDs for the 40$\times$ MO.  We only use a small subset of the LEDs, as detailed in each experiment section.  Different LED sampling patterns are explored and their reconstruction results are compared experimentally, demonstrating the flexibility in data acquisition enabled by our IDT framework.  In all the images, the reconstructed permittivity contrast has been converted to the real (phase) and imaginary (absorption) part of the refractive index.   The background medium index for all the biological samples are assumed to be 1.33 (i.e. close to water).
	
	\subsection{Imaging of stained 3D sample}

	We first demonstrate our technique to image a stained spirogyra sample (Fisher Scientific S68786) using the 10$\times$ MO.  The sample contains both highly absorbing features (e.g. chloroplasts)  and ``phase'' features (e.g. filaments) orientated in 3D, as seen in the full field-of-view (FOV) brightfield image in Fig.~\ref{spiral}(a).  We used all of the 89 brightfield images to perform the IDT reconstruction.  To demonstrate the best possible axial resolution, the slice spacing is set to be $5\mu m$ in the reconstruction, corresponding to approximately 2$\times$ oversampling axially.  We have reconstructed 25 phase and absorption slices equally spaced between $-20\mu m$ and $200\mu m$.   The number of images used is about twice the number of unknowns, the underlying linear problem is thus over-determined, albeit ill-conditioned due to the missing axial frequency information that in turn sets the axial resolution.   
	
	The reconstruction performance is illustrated on features with different length scales.  In Fig.~\ref{spiral}(b), we zoom in on a region containing four clustered spirogyras.   Our  reconstruction demonstrates that axial sectioning can be successfully obtained using our IDT algorithm - features at different depths are clearly distinguished in the reconstruction.   In Fig.~\ref{spiral}(c), we turn to a spiral structure on a single spirogyra.  The helical structure that presents much finer axial features, is successfully reconstructed.  Finally, we look at the filaments in Fig.~\ref{spiral}(d), which is a ``phase sample'' that does not provide high contrast in the brightfield image.   Using our technique, the filaments are clearly resolved with high contrast in the phase reconstruction. As expected, the absorption reconstruction does not provide much contrast.

	\begin{figure*}[!ht]
		\centering
		\includegraphics[width=\linewidth]{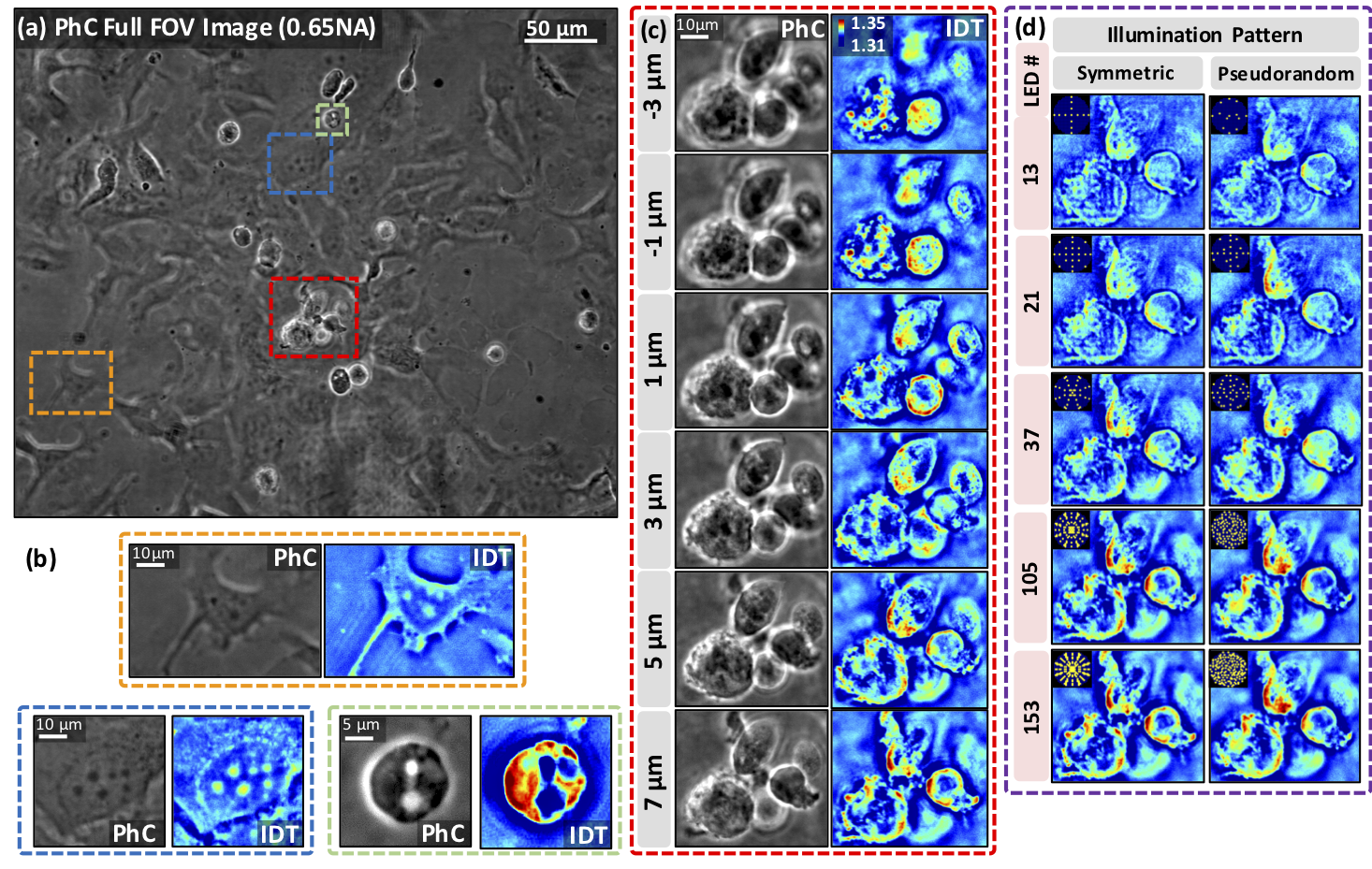}
		\caption{Reconstruction of unstained MCF-7 cancer cells. (a) The full FOV PhC image (40$\times$, 0.65NA).  (b) Phase reconstructions on a few cell regions, demonstrating its versatility and robustness in reconstructing both thin and thick samples.  (c) Phase reconstruction of a dense cell cluster across multiple slices.  The comparison with the physically scanned PhC images demonstrates that our IDT technique provides similar lateral resolution and axial sectioning capability.  (d)  Phase reconstruction of the cell clusters using symmetric and pseudorandom illumination patterns.  Our IDT framework allows flexibly designing the illumination pattern and the number of LEDs used.  The reconstruction algorithm produces high quality phase recovery as the number of images used is reduced, and remains robust even when the number of images is much fewer than the number of unknowns. 
		} 
		\label{MCF}
	\end{figure*} 
	
	\subsection{Imaging of unstained dense cell clusters}
	
	Next, we image unstained MCF-7 cancer cells fixed in formaldehyde solution using the 40$\times$ MO (Fig.~\ref{MCF}).  The sample contains both monolayer cells and dense cell clusters.  A qualitative visualization of the sample is performed via phase contrast (PhC) mode (MO: 0.65 NA, CFI Plan Achro), shown in Fig.~\ref{MCF}(a).  Features imparting longer optical path delay generally produce darker contrast in the PhC image.   Halos are present at the boundaries of thick cell regions.   
	
	We first use 153 images to reconstruct 22 phase and absorption slices with $1\mu m$ spacing from $-6\mu m$ to $15\mu m$.    To demonstrate the versatility of our IDT technique, Fig.~\ref{MCF}(b) shows the phase reconstructions of a few cell regions covering both thin and thick features, along with their PhC images for comparison.  Subcellular features are correctly reconstructed, matching the PhC images.   Features appear darker contrast in PhC are reconstructed with higher refractive index values.  
	
	Next, we zoom-in on a region containing a dense cluster of 5 cells with partial axial overlapping  [Fig.~\ref{MCF}(c)].  Phase reconstructions are shown at 6 slices along with the corresponding PhC images, captured by mechanically adjusting the focus.  Since both techniques use high angle illumination that results in similar Fourier coverage, we expect that they should provide similar lateral resolution and axial sectioning capability. This can be qualitatively verified in Fig.~\ref{MCF}(c).
	
	Finally, we demonstrate the flexibility and robustness of our technique when different illumination patterns and number of LEDs are used for data acquisition.  In Fig.~\ref{MCF}(d), we investigate illumination strategies using symmetric and pseudorandom patterns.  Each symmetric pattern uses LEDs that are equally spaced along both azimuthal and radial directions.   Each pseudorandom pattern is designed such that each quadrant contains the same number of randomly located LEDs.   In all cases, we reconstruct the same 22 phase and absorption slices (i.e. in-total 44 unknown slices).  With 37 images, the number of measurement is slightly smaller than the number of unknowns.  Although the problem is under-determined, the reconstruction only degrades slightly as compared to the ones from using more images (e.g. 105 and 153).   When we further reduce the number of images, the reconstruction further degrades, but is still able to recover cellular features.   Using the pseudorandom pattern does not make significant difference when using a large number of LEDs (e.g. 105 and 153).  As the number of LEDs is reduced, the pseudorandom pattern is observed to achieve slightly better results, likely because it can provide more uniform Fourier coverage.

	\begin{figure*}[!ht]
		\centering
		\includegraphics[width=\linewidth]{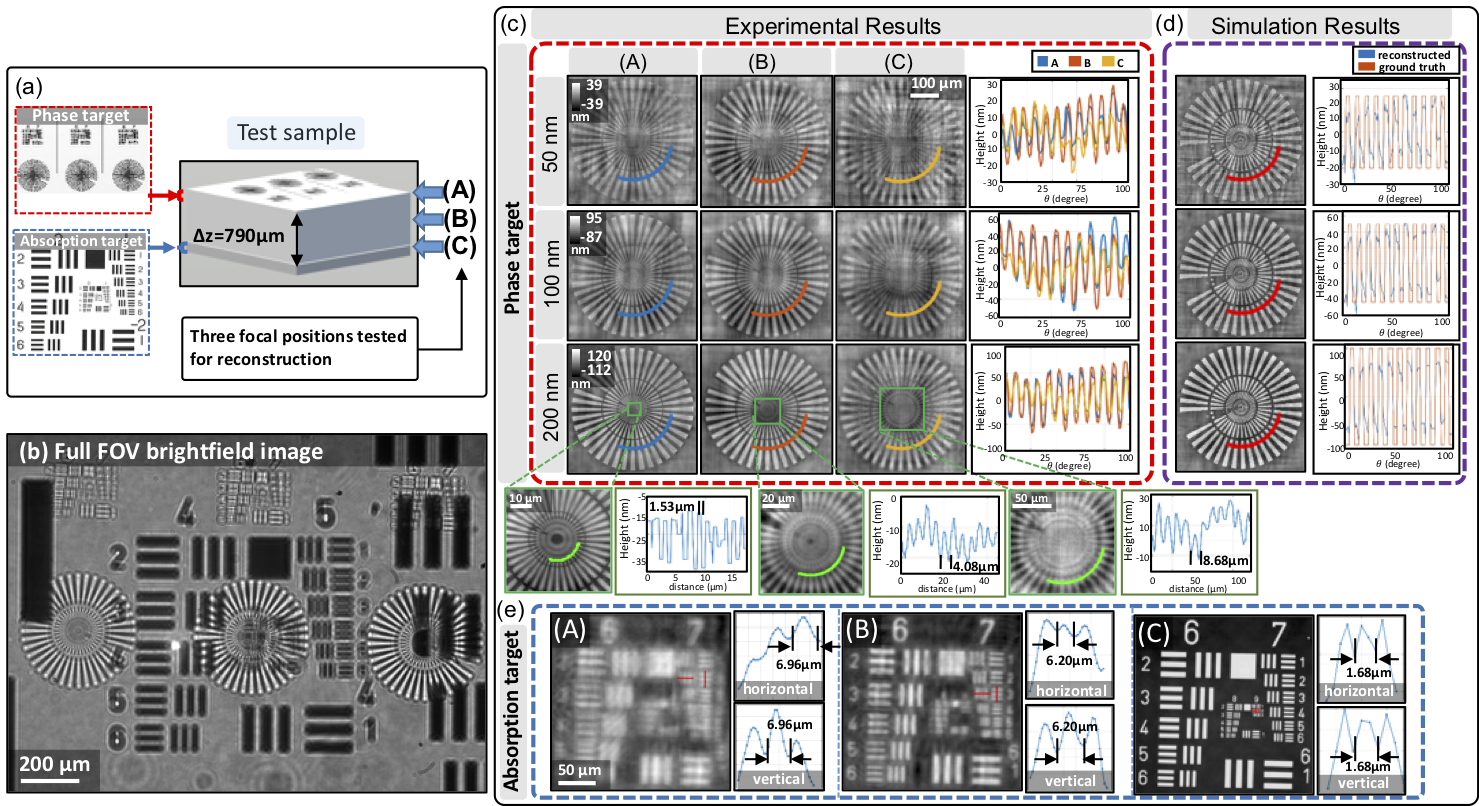}
		\caption{Imaging of strongly scattering phase and absorption targets. (a) The sample consists of a phase target placed above an absorption target.  Experiments are taken at: (A) near the phase target plane, (B) in between the phase and absorption targets, and (C) near the absorption target.  (b) The brightfield image with on-axis LED illumination (10$\times$, 0.25NA).  (c) The reconstruction of phase patterns with 50nm, 100nm, and 200nm in height at the three focal positions.  The reconstruction shows under-estimation of the phase due to multiple scattering.   The reconstruction of the taller pattern contains larger error, because of the stronger multiple scattering.  The recovered resolution when the target is near focus in (A) matches with the theory, and degrades as increasing the defocus.  Nevertheless, the recovered phase values are consistent at all focal positions.  (d) Simulation shows that the unaccounted multiple scattering indeed results in under-estimation of the phase; the amount of reconstruction error matches well with the experiment.  (e) The reconstruction of the absorption target at the three focal positions.  Similar to the phase target, the recovered resolution agrees with the theory when the target is near focus in (C), while degrades with large defocus. 
		}
		\label{target}
	\end{figure*}

	\subsection{Imaging of strongly scattering sample}
	
	Finally, we evaluate our technique when imaging a strongly scattering sample, which contains both highly absorbing and strong phase features with high permittivity contrast.  The goal is to investigate the performance of our technique when multiple scattering effects become strong.  The influence of multiple scattering to ODT has been investigated extensively~\cite{Chen1998, azimi1983distortion}.   A general observation is that the unaccounted-for multiple scattering leads to under-estimation of the permittivity contrast.  The stronger the multiple scattering, the more severe the under-estimation~\cite{Chen1998}.   In IDT, the data is further limited by the intensity-only information.   Our forward model only considers single scattering; intensity variations due to multiple scattering essentially act as {\it non-random} noise in the inversion, so they can only be partially suppressed by the Tikhonov regularization. 
	
	The sample consists of an absorption target (58-198, Edmund Optics) placed underneath a phase target (QPT, Benchmark Technologies), separated axially by $\sim790\mu m$ [Fig.~\ref{target}(a)].  The absorption target contains chromium patterns on a glass substrate.  The phase target consists of structures of different sizes and heights with refractive index 1.52, which are raised above a glass substrate.  All the experiments are performed in air, so the permittivity contrast is large (>1), violating the requirement of the first Born approximation.  To investigate how the reconstruction degrades as the strength of scattering increases, we perform controlled experiments on phase patterns with increasing height (50nm, 100nm, and 200nm), while keeping the same region of the absorption target underneath.  To further investigate the influence of sample depths, we repeat the experiments at three different focal planes.  The first dataset is taken with the phase target in focus [(A) in Fig.~\ref{target}];  the second with focus 300nm below (A) [(B) in Fig.~\ref{target}]; the third with the absorption target in focus [(C) in Fig.~\ref{target}].  In each experiment, we take 89 brightfield images with the 10$\times$ MO.  A sample brightfield image from the on-axis LED is shown in Fig.\ref{target}(b).  
	
	Since the sample consists of only two layers of interest, the reconstruction can be efficiently performed just on the two slices, without the need to compute any slices in-between.  To implement the reconstruction, we use an axial sampling $\Delta z = 10\mu m$, matching the theoretical resolution.  Since $\Delta z$ is much larger than the pattern height, the reconstructed permittivity contrast $\Deps_{\mathrm{rec}}$ should be interpreted as the average over the slice thickness.  Given the refractive index of the phase target $n_{\mathrm{ph}}$, we can convert the averaged permittivity contrast to an estimate of the pattern height by $h_{\mathrm{rec}}=\Deps_{\mathrm{rec}}\Delta z/(n^2_{\mathrm{ph}}-1)$.  The reconstructed pattern heights for the three different cases are shown in Fig.~\ref{target}(c).  Due to the intensity-only measurement, the phase's DC component is always lost, and thus all the height estimates are centered around zero.  Under-estimation of the heights is observed in all cases due to the strong multiple scattering that is unaccounted for in our model.  As we image a taller pattern, the multiple scattering becomes stronger; correspondingly, the reconstruction shows a larger error in the estimation.   The strong multiple scattering also leads to the hazy background in each reconstructed image. 
	
	To quantitatively validate these results, we perform simulations on a similar two-layer sample [Fig.~\ref{target}(d)].   The multislice model~\cite{Tian.Waller2015} is used to simulate the intensity measurements since it can accurately model multiple scattering  at this length scale~\cite{kamilov2016optical}.  We then use our IDT model to perform the reconstruction.  As seen in the recovered images, the amount of height under-estimation in each case matches well with the corresponding experimental results.   A similar level of background noise is also observed.   
	
	Using our IDT model, we expect to achieve the same lateral resolution at all slices, set by the bandwidth ${4\mr{NA}}/{\lambda}$.  As shown in cases (A), when the phase target is near focus, we successfully achieve the theoretically predicted resolution (1.54$\mu m$ based on the Rayleigh Criterion).  However, as the focal position is defocused from the target, the resolution degrades, [(B) and (C) in Fig.~\ref{target}(c)].  The same trend is observed for the absorption target [Fig.~\ref{target}(e)].  A similar resolution degradation was also observed in~\cite{Tian.Waller2015} using a similar setup, though the experiments reported here span $\sim8\times$ wider defocus range (e.g. 800$\mu m$ here vs $110\mu m$ in \cite{Tian.Waller2015}).  We expect that this degradation is likely due to LED position mis-calibration.  The amount of geometrical shift  scales linearly with the defocus distance due to the linear phase term in the TFs.  Thus, the same amount of angular mis-calibration would result in a larger error in data with farther defocus.  Notably, despite this resolution degradation, the recovered phase values remain consistent regardless of the focal positions, as shown in the cutlines of the estimated heights in Fig.~\ref{target}(c).

	\section{Conclusion}
	
	We have demonstrated a new computational microscopy technique that enables scan-free, 3D phase and absorption reconstruction using intensity-only measurements.  Our slice-based intensity diffraction tomography framework allows flexible illumination patterning for data acquisition and a direct inversion algorithm that has both low computational complexity and is memory efficient.  We have experimentally validated this technique on dense 3D biological samples.  Although our model only accounts for single scattering, the reconstruction is robust for samples with high permittivity contrast.  The effects of multiple scattering are examined via both simulation and experiment, and their results have been found to be in good agreement.  The system is simple, and built directly on a commercial microscope with an LED array to enable rapid illumination-angle scanning.  This is particularly attractive for the wide adoption of our system to existing microscopy facilities and may open up many biomedical imaging applications.
	
	\section*{Funding}
	National Science Foundation (NSF) Industry/University Cooperative Research Center for Biophotonic Sensors and Systems (IIP-1068070).
	
	\section*{Acknowledgments}
	
	The authors thank Ziji Liu for the help with the LED array.
	
	\section*{Disclosures}
	The authors declare that there are no conflicts of interest related to this article.


\begin{thebibliography}{10}
	
	\bibitem{Stephens2003}
	D.~J. Stephens and V.~J. Allan.
	\newblock Light microscopy techniques for live cell imaging.
	\newblock {\em Science}, 300(5616):82--86, 2003.
	
	\bibitem{Waeldchen2015}
	S.~W{\"a}ldchen, J.~Lehmann, T.~Klein, S.~Van De~Linde, and M.~Sauer.
	\newblock Light-induced cell damage in live-cell super-resolution microscopy.
	\newblock {\em {Scientific Reports}}, 5, 2015.
	
	\bibitem{Hoebe2007}
	R.~Hoebe, C.~Van~Oven, T.~J. Gadella, P.~Dhonukshe, C.~Van~Noorden, and
	E.~Manders.
	\newblock Controlled light-exposure microscopy reduces photobleaching and
	phototoxicity in fluorescence live-cell imaging.
	\newblock {\em {Nature Biotechnology}}, 25(2):249--253, 2007.
	
	\bibitem{Zheng2011}
	G.~Zheng, C.~Kolner, and C.~Yang.
	\newblock Microscopy refocusing and dark-field imaging by using a simple {LED}
	array.
	\newblock {\em Optics {L}etters}, 36(20):3987--3989, 2011.
	
	\bibitem{Zheng2013}
	G.~Zheng, R.~Horstmeyer, and C.~Yang.
	\newblock Wide-field, high-resolution {Fourier} {P}tychographic microscopy.
	\newblock {\em Nat. Photonics}, 7(9):739--745, 2013.
	
	\bibitem{Tian2014}
	L.~Tian, X.~Li, K.~Ramchandran, and L.~Waller.
	\newblock Multiplexed coded illumination for {Fourier} ptychography with an
	{LED} array microscope.
	\newblock {\em Biomed. Opt. Express}, 5(7):2376--2389, Jul 2014.
	
	\bibitem{Tian2014a}
	L.~Tian, J.~Wang, and L.~Waller.
	\newblock {3D} differential phase-contrast microscopy with computational
	illumination using an {LED} array.
	\newblock {\em Opt. Lett.}, 39(5):1326--1329, Mar. 2014.
	
	\bibitem{Tian.Waller2015}
	L.~Tian and L.~Waller.
	\newblock {3D} intensity and phase imaging from light field measurements in an
	{LED} array microscope.
	\newblock {\em Optica}, 2:104--111, 2015.
	
	\bibitem{horstmeyer2016diffraction}
	R.~Horstmeyer, J.~Chung, X.~Ou, G.~Zheng, and C.~Yang.
	\newblock Diffraction tomography with fourier ptychography.
	\newblock {\em Optica}, 3(8):827--835, 2016.
	
	\bibitem{Tian2015a}
	L.~Tian and L.~Waller.
	\newblock Quantitative differential phase contrast imaging in an {LED} array
	microscope.
	\newblock {\em Opt. Express}, 23(9):11394--11403, May 2015.
	
	\bibitem{Tian2015b}
	L.~Tian, Z.~Liu, L.-H. Yeh, M.~Chen, J.~Zhong, and L.~Waller.
	\newblock Computational illumination for high-speed in vitro {Fourier}
	ptychographic microscopy.
	\newblock {\em Optica}, 2(10):904--911, 2015.
	
	\bibitem{Choi.etal2007}
	W.~Choi, C.~Fang-Yen, K.~Badizadegan, S.~Oh, N.~Lue, R.~R. Dasari, and M.~S.
	Feld.
	\newblock Tomographic phase microscopy.
	\newblock {\em Nat. Methods}, 4(9):717--719, September 2007.
	
	\bibitem{Sung.etal2009}
	Y.~Sung, W.~Choi, C.~Fang-Yen, K.~Badizadegan, R.~R. Dasari, and M.~S. Feld.
	\newblock Optical diffraction tomography for high resolution live cell imaging.
	\newblock {\em Opt. Express}, 17(1):266--277, December 2009.
	
	\bibitem{cotte2013marker}
	Y.~Cotte, F.~Toy, P.~Jourdain, N.~Pavillon, D.~Boss, P.~Magistretti,
	P.~Marquet, and C.~Depeursinge.
	\newblock Marker-free phase nanoscopy.
	\newblock {\em Nature Photonics}, 2013.
	
	\bibitem{kamilov2015learning}
	U.~S. Kamilov, I.~N. Papadopoulos, M.~H. Shoreh, A.~Goy, C.~Vonesch, M.~Unser,
	and D.~Psaltis.
	\newblock Learning approach to optical tomography.
	\newblock {\em Optica}, 2(6):517--522, 2015.
	
	\bibitem{simon2017tomographic}
	B.~Simon, M.~Debailleul, M.~Houkal, C.~Ecoffet, J.~Bailleul, J.~Lambert,
	A.~Spangenberg, H.~Liu, O.~Soppera, and O.~Haeberl{\'e}.
	\newblock Tomographic diffractive microscopy with isotropic resolution.
	\newblock {\em Optica}, 4(4):460--463, 2017.
	
	\bibitem{Charriere:06}
	F.~Charri\`{e}re, A.~Marian, F.~Montfort, J.~Kuehn, T.~Colomb, E.~Cuche,
	P.~Marquet, and C.~Depeursinge.
	\newblock Cell refractive index tomography by digital holographic microscopy.
	\newblock {\em Opt. Lett.}, 31(2):178--180, Jan. 2006.
	
	\bibitem{Zhang:2016aa}
	T.~Zhang, C.~Godavarthi, P.~C. Chaumet, G.~Maire, H.~Giovannini, A.~Talneau,
	M.~Allain, K.~Belkebir, and A.~Sentenac.
	\newblock Far-field diffraction microscopy at $\lambda$/10 resolution.
	\newblock {\em Optica}, 3(6):609--612, 2016.
	
	\bibitem{Bon2014a}
	P.~Bon, S.~Aknoun, S.~Monneret, and B.~Wattellier.
	\newblock Enhanced {3D} spatial resolution in quantitative phase microscopy
	using spatially incoherent illumination.
	\newblock {\em {Optics Express}}, 22(7):8654--8671, 2014.
	
	\bibitem{Kim.etal2014}
	T.~Kim, R.~Zhou, M.~Mir, S.~Babacan, P.~Carney, L.~Goddard, and G.~Popescu.
	\newblock White-light diffraction tomography of unlabelled live cells.
	\newblock {\em Nat. Photonics}, 8:256--263, March 2014.
	
	\bibitem{nguyen2017gradient}
	T.~H. Nguyen, M.~E. Kandel, M.~Rubessa, M.~B. Wheeler, and G.~Popescu.
	\newblock Gradient light interference microscopy for {3D} imaging of unlabeled
	specimens.
	\newblock {\em {Nature Communications}}, 8(1):210, 2017.
	
	\bibitem{kim2014common}
	Y.~Kim, H.~Shim, K.~Kim, H.~Park, J.~H. Heo, J.~Yoon, C.~Choi, S.~Jang, and
	Y.~Park.
	\newblock Common-path diffraction optical tomography for investigation of
	three-dimensional structures and dynamics of biological cells.
	\newblock {\em {Optics Express}}, 22(9):10398--10407, 2014.
	
	\bibitem{kus2014tomographic}
	A.~Ku{\'s}, M.~Dudek, B.~Kemper, M.~Kujawi{\'n}ska, and A.~Vollmer.
	\newblock Tomographic phase microscopy of living three-dimensional cell
	cultures.
	\newblock {\em {Journal of Biomedical Optics}}, 19(4):046009--046009, 2014.
	
	\bibitem{habaza2015tomographic}
	M.~Habaza, B.~Gilboa, Y.~Roichman, and N.~T. Shaked.
	\newblock Tomographic phase microscopy with 180 rotation of live cells in
	suspension by holographic optical tweezers.
	\newblock {\em Optics {L}etters}, 40(8):1881--1884, 2015.
	
	\bibitem{kim2015simultaneous}
	K.~Kim, J.~Yoon, and Y.~Park.
	\newblock Simultaneous {3D} visualization and position tracking of optically
	trapped particles using optical diffraction tomography.
	\newblock {\em Optica}, 2(4):343--346, 2015.
	
	\bibitem{kim2017tomographic}
	K.~Kim and Y.~Park.
	\newblock Tomographic active optical trapping of arbitrarily shaped objects by
	exploiting 3d refractive index maps.
	\newblock {\em {Nature Communications}}, 8:15340, 2017.
	
	\bibitem{shin2015active}
	S.~Shin, K.~Kim, J.~Yoon, and Y.~Park.
	\newblock Active illumination using a digital micromirror device for
	quantitative phase imaging.
	\newblock {\em Optics {L}etters}, 40(22):5407--5410, 2015.
	
	\bibitem{Maleki:93}
	M.~H. Maleki and A.~J. Devaney.
	\newblock Phase-retrieval and intensity-only reconstruction algorithms for
	optical diffraction tomography.
	\newblock {\em J. Opt. Soc. Am. A}, 10(5):1086--1092, May 1993.
	
	\bibitem{Gbur:02ol}
	G.~Gbur and E.~Wolf.
	\newblock Diffraction tomography without phase information.
	\newblock {\em Opt. Lett.}, 27(21):1890--1892, Nov. 2002.
	
	\bibitem{Gbur:02}
	G.~Gbur and E.~Wolf.
	\newblock Hybrid diffraction tomography without phase information.
	\newblock {\em J. Opt. Soc. Am. A}, 19(11):2194--2202, Nov. 2002.
	
	\bibitem{Gbur:05}
	G.~Gbur, M.~A. Anastasio, Y.~Huang, and D.~Shi.
	\newblock Spherical-wave intensity diffraction tomography.
	\newblock {\em J. Opt. Soc. Am. A}, 22(2):230--238, Feb. 2005.
	
	\bibitem{Anastasio:05}
	M.~A. Anastasio, D.~Shi, Y.~Huang, and G.~Gbur.
	\newblock Image reconstruction in spherical-wave intensity diffraction
	tomography.
	\newblock {\em J. Opt. Soc. Am. A}, 22(12):2651--2661, Dec. 2005.
	
	\bibitem{Jenkins2015}
	M.~H. Jenkins and T.~K. Gaylord.
	\newblock Three-dimensional quantitative phase imaging via tomographic
	deconvolution phase microscopy.
	\newblock {\em {Applied Optics}}, 54(31):9213--9227, 2015.
	
	\bibitem{Soto2017}
	J.~M. Soto, J.~A. Rodrigo, and T.~Alieva.
	\newblock Label-free quantitative 3d tomographic imaging for partially coherent
	light microscopy.
	\newblock {\em Optics Express}, 25(14):15699--15712, 2017.
	
	\bibitem{Rodrigo2017}
	J.~A. Rodrigo, J.~M. Soto, and T.~Alieva.
	\newblock Fast label-free microscopy technique for 3d dynamic quantitative
	imaging of living cells.
	\newblock {\em Biomedical Optics Express}, 8(12):5507--5517, 2017.
	
	\bibitem{soto2018optical}
	J.~M. Soto, J.~A. Rodrigo, and T.~Alieva.
	\newblock Optical diffraction tomography with fully and partially coherent
	illumination in high numerical aperture label-free microscopy.
	\newblock {\em Applied Optics}, 57(1):A205--A214, 2018.
	
	\bibitem{Chen2016}
	M.~Chen, L.~Tian, and L.~Waller.
	\newblock {3D differential phase contrast microscopy}.
	\newblock {\em Biomed. Opt. Express}, 7(10):3940--3950, Oct 2016.
	
	\bibitem{N19846}
	N.~Streibl.
	\newblock Phase imaging by the transport equation of intensity.
	\newblock {\em Opt. Commun.}, 49(1):6--10, 1984.
	
	\bibitem{Wolf1969}
	E.~Wolf.
	\newblock Three-dimensional structure determination of semi-transparent objects
	from holographic data.
	\newblock {\em Opt. Commun.}, 1(4):153--156, September/October 1969.
	
	\bibitem{Levoy2006}
	M.~Levoy, R.~Ng, A.~Adams, M.~Footer, and M.~Horowitz.
	\newblock Light field microscopy.
	\newblock In {\em SIGGRAPH '06}, pages 924--934, 2006.
	
	\bibitem{Mehta2009}
	S.~Mehta and C.~Sheppard.
	\newblock Quantitative phase-gradient imaging at high resolution with
	asymmetric illumination-based differential phase contrast.
	\newblock {\em Opt. {L}ett.}, 34(13):1924--1926, 2009.
	
	\bibitem{Born1999}
	M.~{Born} and E.~{Wolf}.
	\newblock {\em {Principles of Optics: Electromagnetic Theory of Propagation,
			Interference and Diffraction of Light}}.
	\newblock Cambridge University Press, 7 edition, Oct. 1999.
	
	\bibitem{Streibl:85}
	N.~Streibl.
	\newblock Three-dimensional imaging by a microscope.
	\newblock {\em J. Opt. Soc. Am. A}, 2(2):121--127, 1985.
	
	\bibitem{Sheppard1994}
	C.~J.~R. Sheppard, Y.~Kawata, S.~Kawata, and M.~Gu.
	\newblock Three-dimensional transfer functions for high-aperture systems.
	\newblock {\em J. Opt. Soc. Am. A}, 11(2):593--598, Feb 1994.
	
	\bibitem{lue2008synthetic}
	N.~Lue, W.~Choi, G.~Popescu, K.~Badizadegan, R.~R. Dasari, and M.~S. Feld.
	\newblock Synthetic aperture tomographic phase microscopy for {3D} imaging of
	live cells in translational motion.
	\newblock {\em {Optics Express}}, 16(20):16240--16246, 2008.
	
	\bibitem{lee2013synthetic}
	K.~Lee, H.-D. Kim, K.~Kim, Y.~Kim, T.~R. Hillman, B.~Min, and Y.~Park.
	\newblock Synthetic {Fourier} transform light scattering.
	\newblock {\em {Optics Express}}, 21(19):22453--22463, 2013.
	
	\bibitem{bertero1998introduction}
	M.~Bertero and P.~Boccacci.
	\newblock {\em Introduction to inverse problems in imaging}.
	\newblock Taylor \& Francis, 1998.
	
	\bibitem{Chen1998}
	B.~Chen and J.~J. Stamnes.
	\newblock Validity of diffraction tomography based on the first {Born and the
		first Rytov} approximations.
	\newblock {\em Applied Optics}, 37(14):2996--3006, 1998.
	
	\bibitem{azimi1983distortion}
	M.~Azimi and A.~Kak.
	\newblock Distortion in diffraction tomography caused by multiple scattering.
	\newblock {\em IEEE Trans. Med. Imaging}, 2(4):176--195, 1983.
	
	\bibitem{kamilov2016optical}
	U.~S. Kamilov, I.~N. Papadopoulos, M.~H. Shoreh, A.~Goy, C.~Vonesch, M.~Unser,
	and D.~Psaltis.
	\newblock Optical tomographic image reconstruction based on beam propagation
	and sparse regularization.
	\newblock {\em IEEE Transactions on Computational Imaging}, 2(1):59--70, 2016.

\end{thebibliography}
	

\end{document}